\documentclass[english,aps,prl,preprint, showpacs, groupedaddress, amsmath, keywords, showpacs,floatfix]{revtex4}
\usepackage{ulem}
\usepackage{fancyhdr,graphicx}
\usepackage{bm}
\usepackage{color}
\usepackage{amsfonts}
\usepackage{times}
\usepackage{amsmath}
\usepackage{parskip} 
\usepackage{natbib}
\usepackage{chemformula} 
\usepackage[colorlinks=true, citecolor=red]{hyperref}
\setcitestyle{journalcolor= blue}
\usepackage{chemformula} 
\usepackage[T1]{fontenc} 

\usepackage{amsmath} 

\usepackage{pdfpages}      

\usepackage{babel}

\begin{document}
\title{Electric field tunable bands in doubly aligned bilayer graphene hBN moir\'{e} superlattice}
	\author{Priya Tiwari$^{1,2,\textcolor{red}{*}}$, Kenji Watanabe$^3$, Takashi Taniguchi$^4$, Aveek Bid$^1$}	\email{priya.tiwari@weizmann.ac.il}
        \email{aveek@iisc.ac.in}
	\affiliation{$^1$Department of Physics, Indian Institute of Science, Bangalore 560012, India \\
  $^2$ Braun Center for Submicron Research, Department of Condensed Matter Physics, Weizmann Institute of Science, Rehovot 76100, Israel\\
	$^3$ Research Center for Functional Materials, National Institute for Materials Science, 1-1 Namiki, Tsukuba 305-0044, Japan \\
	$^4$ International Center for Materials Nanoarchitectonics, National Institute for Materials Science, 1-1 Namiki, Tsukuba 305-0044, Japan}
		 
\begin{abstract}
		
In this letter, we demonstrate electric field-induced band modification of an asymmetrically twisted hBN/BLG/hBN supermoir\'e lattice. Distinct from unaligned BLG/hBN systems, we observe regions in the density-displacement field $(n-D)$ plane where the device conductance is independent of $n$ and decreases as $|D|$ increases. This distinction arises due to the angle asymmetry between the layers, which induces field-controlled layer polarization. We identify $D$-dependent additional band gaps near the charge neutrality point that appear in the conduction (valence) band for negative (positive) $D$ values. In the quantum Hall regime, new $6$-fold degenerate Landau levels are observed. Our findings establish that in an asymmetric supermoir\'e heterostructure, an external vertical displacement field affects the valence and conduction bands very differently and sheds light on the asymmetric conductance patterns noted in previous studies. 

\end{abstract}
\maketitle	

\section{Introduction}

Heterostructures of two-dimensional van der Waals materials are pivotal in probing new quantum phases. Stacking two-dimensional heterostructures allows the merging of diverse materials with unique properties, potentially creating a broad platform for exploring quantum phenomena~\cite{Liu2016, Geim2013,doi:10.1146/annurev-matsci-070214-020934, C4NR03435J,doi:10.1002/adfm.201706587, Novoselovaac9439,tiwari2020observation,tiwari2021electric,tiwari2023observation,doi:10.1021/acs.nanolett.3c04223}.
Till recently, the prevailing best practice was to produce heterostructures of multiple two-dimensional materials with complementary properties. However, recent research has revealed surprising findings in stacks made by repeatedly layering one or two ultra-thin materials with a slight twist angle between the crystallographic axis of each subsequent layer~\cite{cao2018unconventional,cao2020tunable,kazmierczak2021strain,li2024tuning, cai2023signatures,xu2023observation,jat2023higher, scuri2020electrically,vitale2021flat,li2024tuning}.

A sheet of single-layer graphene (SLG) or bilayer graphene (BLG) encased by two slightly misaligned thin flakes of hexagonal boron nitride (hBN) provides one such exciting environment~\cite{gonzalez2021topological,song2015topological,ponomarenko2013cloning,Wang2015,doi:10.1063/1.5094456,Chen2020, jat2023higher}. The mismatch in lattice constants between BLG and hBN, along with the angular misalignment of the layers, leads to the formation of two distinct, moir\'e patterns at the top and bottom graphene-hBN interfaces~\cite{Yankowitz2012, Yankowitz2016,doi:10.1126/sciadv.abd3655,Finney2019,doi:10.1021/acs.nanolett.8b03423}. The interference between these patterns produces a so-called supermoir\'e structure characterized by complex periodicity in real space~\cite{PhysRevB.103.075122,doi:10.1021/acs.nanolett.9b04058, PhysRevB.104.035306,Leconte_2020, PhysRevB.102.045409,doi:10.1126/sciadv.aay8897,doi:10.1126/sciadv.aay8897,sun2021, PhysRevB.103.115419, Yankowitz2019,Moriya2020}. The large supermoir\'e length scale ($\sim 35$~nm) effectively compresses the graphene band structure into a smaller Brillouin zone~\cite{Mayo_2020} producing additional moiré-induced gaps in the energy spectrum~\cite{jat2023higher}. Recent insights reveal that these superlattice effects also lead to more subtle influences on graphene's electronic properties~\cite{doi:10.1021/acs.nanolett.3c04223}, notably flat bands in the energy dispersion at low energies~\cite{zhu2022electric,doi:10.1021/acs.nanolett.9b04058}. 

The physics becomes even more fascinating if the twist angles between the top hBN and the BLG ($\theta_t$) and between the bottom hBN and BLG ($\theta_b$) are different. This asymmetry breaks the inversion symmetry of the device, leading to layer polarization and creating an inbuilt intrinsic electric field $E_i$ perpendicular to the plane of the device. 
A recent theoretical study finds that in this state, the orbitals forming the valence (conduction) band are predominantly localized in the layer of BLG that forms the larger (smaller) twist angle with hBN~\cite{zhu2022electric}. 
The application of an external electric field ${E}_{ext}$ perpendicular to the plane of such a supermoir\'e device provides an additional degree of freedom that can change this band arrangement drastically~\cite{zhu2022electric, tiwari2021electric, PhysRevLett.119.146401,island2019spin}. 
For ${E}_{ext}$ in the direction opposite to ${E}_i$, and $|{E}_{ext}|>|{E}_i|$, the bands flip, with the valence (conduction) band now layer polarized to the layer forming the smaller (larger) twist angle. This band rearrangement is predicted to lead to additional band gaps and the suppression of the density of states~\cite{zhu2022electric}. These changes should be reflected in electrical transport measurements, leading to the intriguing possibility of an displacement-field controlled supermoir\'e valve.

In this letter, we present the results of a detailed experimental study of the effect of displacement field on the transport properties of hBN/BLG/hBN supermoir\'e heterostructures with $\theta_t \neq \theta_b$. We observe non-trivial dependence of the conductance on an externally applied displacement field $D = E_{ext}/\epsilon$; these findings match very well with theoretical predictions of the variation of the density of states (DOS) with the displacement field~\cite{zhu2022electric}. We find that depending on the relative orientations of ${E}_{ext}$ and ${E}_{i}$, either the electron conductance or the hole conductance can be suppressed by two orders of magnitude while leaving the conductance of the other charge carrier species almost unchanged. 
Additionally, we observe preliminary signatures of quasicrystal formation in the presence of finite $D$ field, as recently observed in twisted trilayer graphene~\cite{Aviram@2023nature, hao2024robust}. 

\section{Results and Discussion}
\begin{figure}[t]
	\includegraphics[width=1\columnwidth]{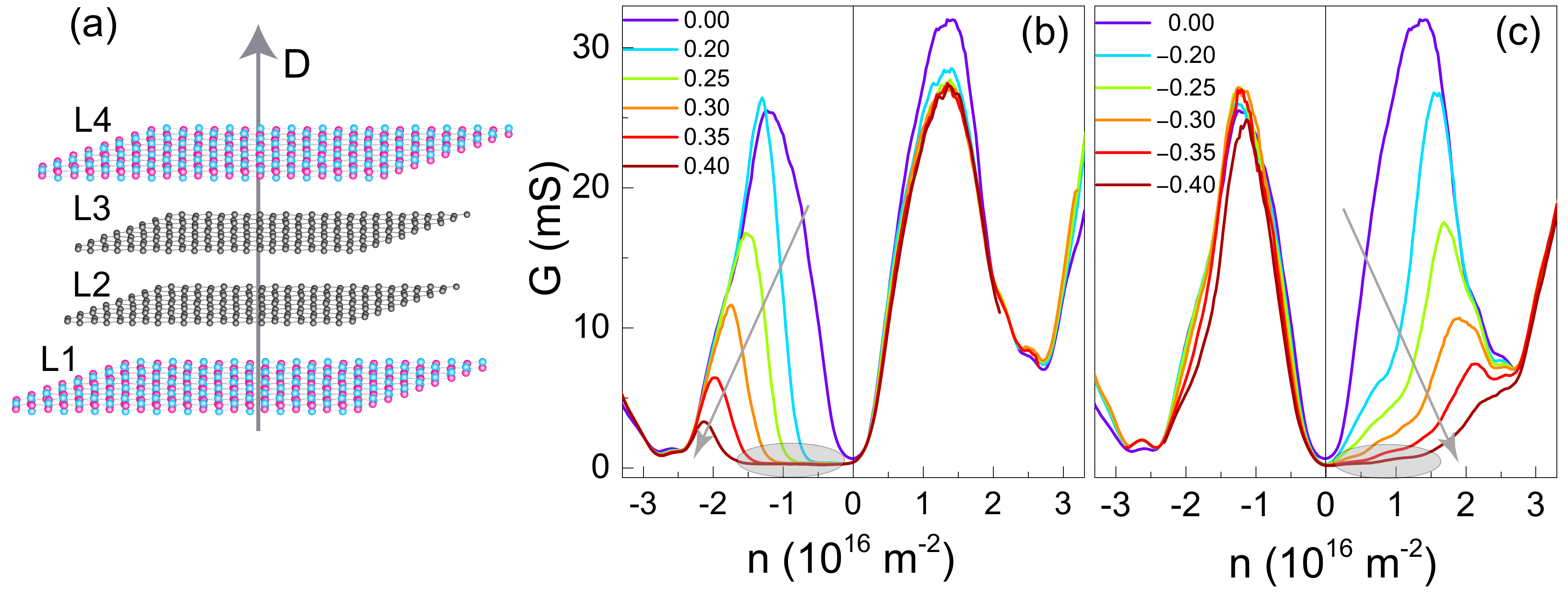}
	\small{\caption{(a) A Schematic of the BLG layer sandwiched between two hBN layers. L1, L2, L3, and L4 are the bottom hBN, the bottom layer of BLG graphene, the top layer of BLG graphene, and the top hBN surfaces. The arrow marks the direction of positive $D$. (b), (c) ${G} (B=0)$ as a function of $n$ for different values of $D$. The labels in the plots indicate the value of $D/\epsilon_0$ in units of V/nm. 
 \label{fig:fig1}}}
\end{figure}

Heterostructures of BLG doubly aligned with hBN with a twist angle of less than $1^\circ$ were fabricated using the dry transfer technique~\cite{pizzocchero2016hot,wang2013one}. These were patterned into dual-gated Hall bars with Cr/Au one-dimensional electrical contacts. This device structure allows independent control of the charge carrier density $n$ and displacement field perpendicular to the BLG plane, $D$ through the relations $n=(({C}_{bg}{V}_{bg}+{C}_{tg}{V}_{tg})/e)+n_0$ and $D=({C}_{bg}{V}_{bg}-{C}_{tg}{V}_{tg})/2+D_0$. Here, $n_0$ is the residual charge density due to unintentional channel doping, $D_0$ is the net internal displacement field, and $V_{tg}$ ($V_{bg}$) is the top-gate (back-gate) voltage. The top and bottom gate capacitance ${C}_{tg},{C}_{bg}$ are extracted from quantum hall measurements. All electrical measurements were performed using low-frequency AC lock-in techniques ($i = 10$~nA, frequency $f = 17$~Hz) at T$\sim2$~K in a $^4$He refrigerator unless otherwise specified. 
As detailed in a previous study on this device, the variation of the four terminal longitudinal resistance $\mathrm{R}_{xx}$ with the number density $n$ showed doubly-split secondary Dirac points at $\pm n_t$ and $\pm n_b$, in addition to the peak at the primary Dirac point ($n=0$), indicating a double alignment of the BLG with both the top and bottom hBN layers~\cite{jat2023higher}. 
From measurements of Brown Zak oscillations, the twist angles are found to be $\theta_b = 0.03^\circ$ (between bottom hBN and BLG) and $\theta_t = 0.44^\circ$ (between top hBN and BLG) with the corresponding moir\'{e} lengths $\lambda_b = 13.97$~nm and $\lambda_t = 12.84$~nm, respectively~\cite{jat2023higher}. 

\subsection{$D$-field induced electron-hole asymmetry }
We now proceed to study the $D$-dependence of the four-probe conductance $G$. Looking at Fig.~\ref{fig:fig1}(b), for $D/\epsilon_0>0$ and $n>0$, we find the conductance to be almost independent of $D$. By contrast, for $D>0$ and $n<0$, with increasing $D$, the conductance peak decreases continuously while steadily shifting towards the secondary Dirac points. It merges with the conductance minima at the secondary Dirac points for $D/\epsilon_0>0.3$~V/nm. An arrow marks this trend in Fig.~\ref{fig:fig1}(b). Concurrently, the conductance near the Dirac point (region indicated by the dotted ellipse) flattens out, dropping by almost two orders of magnitude and approaching zero (Supplementary Information, Fig.~S1 and section~S4). The reverse scenario is played out for negative $D$ (Fig.~\ref{fig:fig1}(c)). For $D<0$ and $n> 0$, $G$ has a strong $D$-dependence while it is almost $D$-independent for $D<0$ and $n<0$. As shown below, the asymmetry demonstrated here results from the combined effects of the asymmetric moi\'e potential across the two layers of BLG and the $D$-induced layer polarization within the BLG. 

From Fig.~\ref{fig:fig1}(b-c), we conclude that for $D>0$ ($D<0$), the effect of the moir\'e potential on the valence (conduction) band is much stronger than on the conduction (valence) band. To understand this, recall that a supermoir\'e structure with $\theta_b \neq\theta_t$ breaks inversion symmetry, leading to an in-built electric field $E_i$ perpendicular to the plane of the device. This field layer polarizes the BLG even without an external displacement field. In our case, $\theta_t >> \theta_b$. Consequently, as discussed in the introduction, for externally applied $D=0$, the wave functions of the carriers in the valence band are predicted to be localized on the interface between the top hBN and BLG (interface of L3/L4 of Fig.~\ref{fig:fig1}(a)). The wave functions of the charge carriers in the conduction band are localized on the interface between the bottom hBN and BLG (interface between L1/L2). This spatial separation of the conduction and valence band carriers exposes the holes in the valence band and electrons in the conduction band to the different moir\'e periodic potentials of the L3/L4 and L1/L2 supercell, respectively.
 \begin{figure}[t]
 	\begin{center}
               \includegraphics[width=0.6\columnwidth]{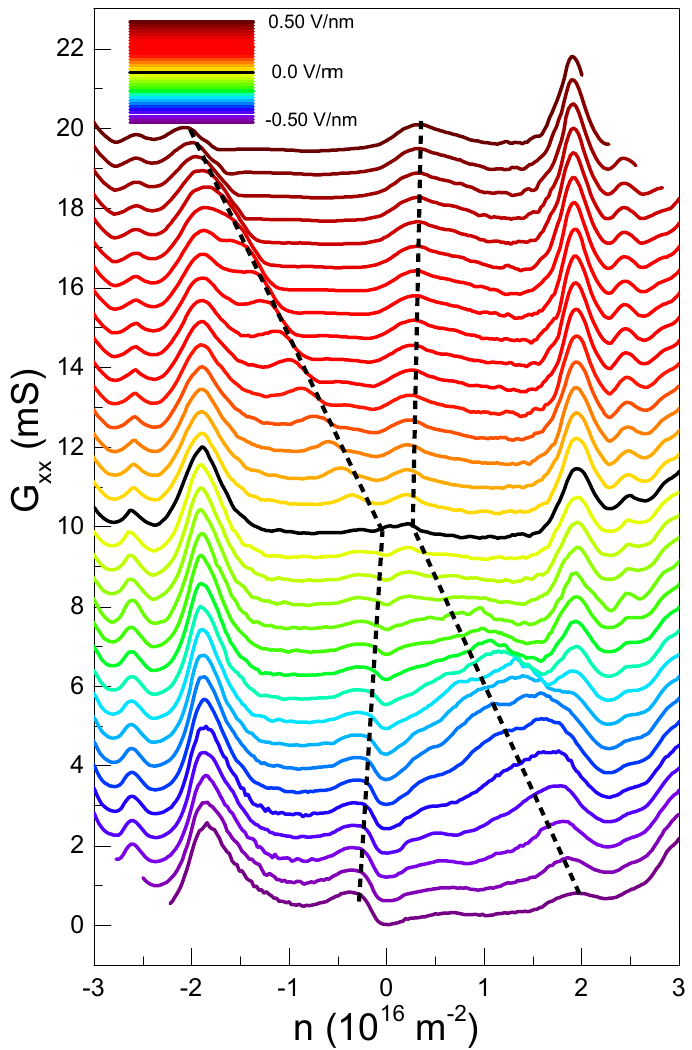} 
 		\small{\caption{Plot of $\mathrm{G}_{xx}$ as a function of $n$ for different values of $D/\epsilon_0$ at $B=0.7$~T. Each curve is vertically shifted by $0.6$~mS for better clarity. The dotted lines mark the evolution of the additional peaks observed at a finite-$D$. }
 			\label{fig:fig2}}
 	\end{center}
 \end{figure}
 
In the presence of an externally applied displacement field $D$, this spatial separation of the carriers in the conduction and the valence bands gets enhanced. Band structure calculations indicate that for increasing positive $D$, the DOS of the conduction band remains largely unaffected, while that for the valence band is strongly suppressed~\cite{zhu2022electric}. Consequently, for large positive $D$, the electron conductance remains almost unchanged, while that for the holes reduces, in conformity with our observations (Fig.~\ref{fig:fig1}(b)). On the other hand, a large negative $D$ will reduce the electron conductance while having a negligible effect on hole conductance, just as seen in Fig.~\ref{fig:fig1}(c).
\begin{figure}[t]
	\begin{center}
                \includegraphics[width=1\columnwidth]{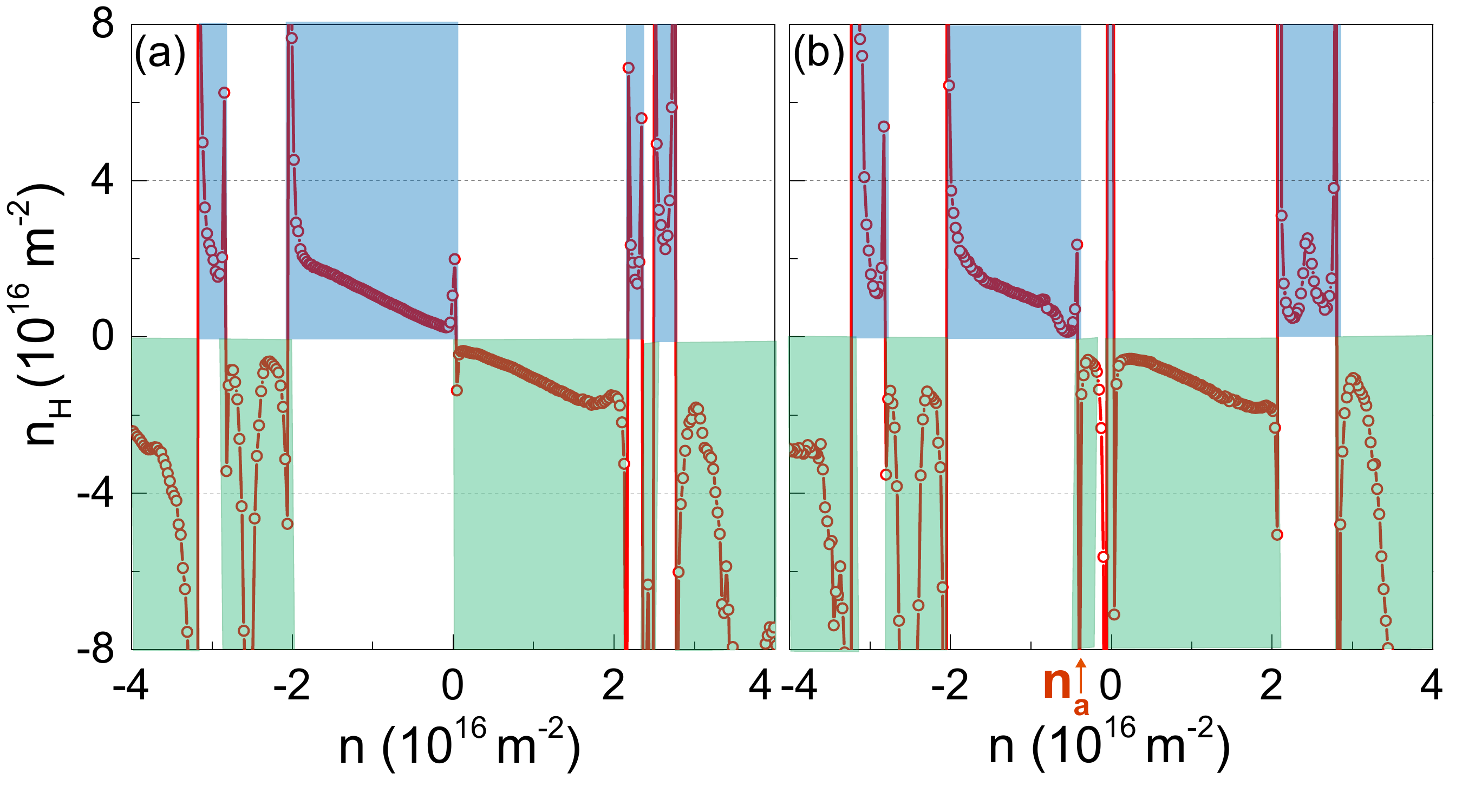}
				\small{\caption{Hall carrier density $n_H$ measured at $B=0.5$~T as a function of number density $n$ for (a) $D/\epsilon_0= 0.0$~V/nm, and (b) $D/\epsilon_0=0.18$~V/nm.}  \label{fig:fig3}}
			
	\end{center}
\end{figure}

In Fig.~\ref{fig:fig2}, we plot the longitudinal conductance $\mathrm{G}_{xx}$ at $B=0.7$~T. Each curve is shifted vertically by $0.6$~mS for clarity. With increasing $|D|$, two additional peaks appear close to the Dirac point on either side of it. The dotted lines mark their evolution with $D$ in Fig.~\ref{fig:fig2}. For $D>0$, the peak for $n>0$ has a shallow dependence on the strength of $D$, while the one for $n<0$ is strongly dispersive. For $D<0$, the opposite happens: the additional peak for $n>0$ is dispersive (and merges with the secondary moir\'e peaks at high $|D|$) while that for peak $n<0$ has a very weak $D$--dependence. These results match qualitatively with the predictions of an asymmetrical dependence of the DOS on $D$~\cite{zhu2022electric}.

\subsection{$D$-field induced modification of gaps}
We now move on to the effect of a $D$ field on the Landau spectrum of the BLG-hBN double-moir\'e system. In Fig.~\ref{fig:fig3} (a-b), we plot the dependence of Hall number density $n_H$ versus the total carrier density $n$ for two representative values of $D$. The values of $n_H$ are extracted from the Hall measurements in a small, non-quantizing magnetic field ${B} = 0.5$~T. Along with the expected change in sign of $n_H$ at the primary and secondary Dirac points, an additional sign change is observed at $n= n_a$ between the primary Dirac point and the secondary Dirac points for $n<0$ at $D/\epsilon_0= 0.18$~V/nm (see Fig.~\ref{fig:fig3} (b)). Below, we establish from quantum Hall measurements that at $n= n_a$, a displacement-field-induced gap opens in the DOS. The data for negative $D$ is shown in~\cite{supplement}.

\begin{figure}[t]
	\begin{center}
 		\includegraphics[width=0.8\columnwidth]{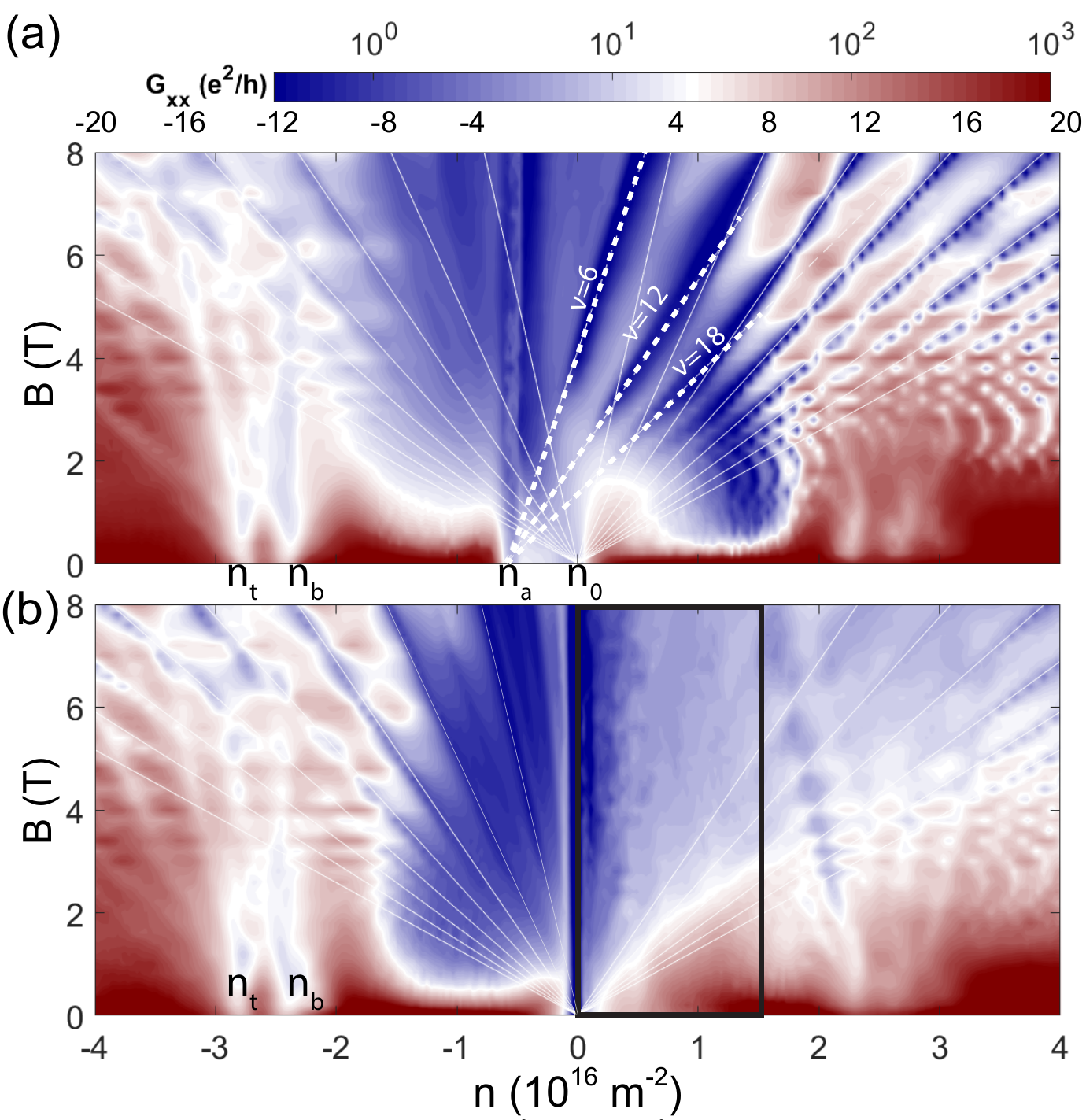}
		\small{\caption{(a)  $\mathrm{G}_{xx}$ as a function of $B$ and $n$ at $D/\epsilon_0=0.18$~V/nm; the solid white lines indicate the conductance minima at $\nu = \pm4,\pm8, \pm12, .....$. An additional landau fan emerges from $n = -0.6 \times 10^{16}$~$ \mathrm{m^{-2}}$ with a six-fold degeneracy (indicated by thick white dotted lines). (b) $\mathrm{G}_{xx}$ as a function of $B$ and $n$ at $D/\epsilon_0=-0.32$~V/nm. the solid white lines indicate the conductance minima at $\nu = \pm4,\pm8, \pm12, .....$. For small positive $n$ between $n=0$ to $n=2\times10^{16}$~m$^{-2}$ the features in G$_{xx}$ is missing. A black rectangle marks the region.}
			\label{fig:fig4}}
	\end{center}
\end{figure}
In Fig.~\ref{fig:fig4}, we plot the Landau fan diagram measured at different values of $D$. For $D/\epsilon_0 = 0$~V/nm, the fan diagram is `conventional' -- one sees Landau fans emanating from the primary Dirac point as well as the secondary Dirac points~\cite{jat2023higher}. The data changes drastically for a finite $D$. For ${D/\epsilon_0}= 0.18$~V/nm (Fig.~\ref{fig:fig4}(a)), we observe an additional fan emerging from $n = -0.6 \times 10^{16}$~$ \mathrm{m^{-2}}$ (marked by thick white dotted line). The vertex of this additional Landau fan coincides with the number density $n_a$ at which $n_H$ shows a sign change. The value of $n_a$ increases with increasing $D$ (see Ref.~\cite{supplement} for a plot of $n_a$ versus $D$). We do not have an understanding of the origin of this $D$-dependent gap. 

Interestingly, this new set of Landau levels is six-fold degenerate, with the minima in $\mathrm{R}_{xx}$ appearing for $\nu=\pm6, \pm12, \pm18$. The appearance of Quantum Hall states with a six-fold degeneracy, as compared to the usual eight-fold degeneracy of pristine BLG, is non-trivial. In a graphene moir\'e system, a twelve-fold degenerate Landau fan is expected at the secondary Dirac cones~\cite{chen2020electrostatic}. Six secondary cones are formed at the \textbf{M} points within each mini-Brillouin zone, with each cone shared by two neighboring mini-Brillouin zones, giving rise to a three-fold degeneracy~\cite{chen2020electrostatic}. Including spin and valley, the expected degeneracy is $3\times 4$. Lifting either spin (due to interactions) or valley (due to breaking of the inversion symmetry) degeneracy will retain an SU(2) symmetry within the still degenerate space and account for the observed $3\times 2$ degeneracy in our system. Further study is required to establish if something similar is the case at $n=n_a$. It should be noted that as we change the displacement field, Landau Fan emerging from the primary Dirac point evolves significantly with displacement field, however the Landau Fan emerging from the moire bands shows a weak dependence on the displacement field. This observation is consistent with the theoretical prediction of Ref.\cite{zhu2022electric}, where it is shown that layer polarization mostly affects the lowest energy bands, and moire bands gets less affected by the layer polarization, and has contribution from both the layers. 

In passing, we note that at high negative values of $D$ and small $n$, the visibility of Landau oscillations reduces. This can be seen, for example, in the region marked by a solid block rectangle in Fig.~\ref{fig:fig4}(b), for $0 < n <2\times10^{16}$~m$^{-2}$ (data for other values of $D$ are shown in~\cite{supplement}). This may be due to the formation of a quasicrystal structure in our double-moir\'e device~\cite{Aviram@2023nature, hao2024robust}. By definition, a quasicrystal has an ordered structure but lacks translational symmetry. It has been proposed that the interference of the two moir\'e structures with different wavelengths can lead to forming a quasicrystal phase~\cite{Aviram@2023nature, hao2024robust}. In such a system, the Landau levels broaden due to increased umklapp scattering. The consequent band reconstruction leads to a reduction in Landau level visibility~\cite{Aviram@2023nature}. Based on the existing theoretical work~\cite{PhysRevB.104.035306} and following the argument of the twisted trilayer graphene work~\cite{Aviram@2023nature}, we provisionally attribute the vanishing visibility to the emergence of the quasicrystalline phase. We hope that our data will motivate future studies on this.

\section{Conclusions} 
We have experimentally demonstrated the displacement-field-induced modification of the energy spectrum in an asymmetrically twisted hBN/BLG/hBN supermoir\'{e} lattice. Our study establishes that in the presence of an external vertical displacement field, the valence band and conduction band encounter different superlattice potentials, and these can be flipped by using the external displacement field as a switch. Specifically, we observe that the angle asymmetry between the top and bottom layers induces layer polarization and significantly alters transport properties, as reflected in resistance features that are asymmetric between conduction and valence bands and tunable by displacement field \( D \). These findings qualitatively align with theoretical predictions [Phys. Rev. B 106, 205134 (2022)] and explains the asymmetric features seen in the conductance map in the $n-D$ plane in previous studies~\cite{jat2023higher, Zheng2020}.  Additionally, a gap emerges at a finite displacement field and a density $(n_a)$ near charge neutrality, accompanied by a distinct Landau fan -- these features had escaped observation in prior studies on supermoir\'e systems. The new gap exhibits prominent Landau oscillations akin to primary and secondary Dirac points, although its origin remains unclear. Interestingly, $n_a$ increases with $D$, and the Landau level degeneracy at this density is $6$, contrasting with the expected $4$ in graphene. Furthermore, we observe reduced visibility in Landau fans at large displacement fields. The quasi-periodicity of hBN/graphene/hBN supermoiré structures and theoretical predictions suggest a strong connection to quasicrystal formation, supported by recent experimental work on twisted trilayer graphene, which attributes similar reductions in Landau oscillation visibility to quasicrystalline phases~\cite{PhysRevB.104.035306, Aviram@2023nature}. 

Our study raises several unanswered questions on the effect of a displacement field on the transport properties of supermoir\'e structures, the most prominent of these being: (1) the origin of the additional band gaps at $n = n_a$; and (2) the effect of the moir\'e lattice strength on the visibility of the quantum Hall fan diagram. We leave these for queries for future studies.

\section{Acknowledgment} 
P.T. acknowledges fruitful discussions with Saurabh Kumar Srivastav. A.B. acknowledges funding from U.S. Army DEVCOM Indo-Pacific (Project number: FA5209   22P0166) and Department of Science and Technology, Govt of India (DST/SJF/PSA-01/2016-17). K.W. and T.T. acknowledge support from the JSPS KAKENHI (Grant Numbers 21H05233 and 23H02052) and World Premier International Research Center Initiative (WPI), MEXT, Japan.
\clearpage
%
\onecolumngrid
\newpage
\thispagestyle{empty}
\mbox{}


\section*{Supplementary material}

## S1. EFFECT OF $D$ ON THE RESISTANCE PEAKS AT SDP

The longitudinal conductance as a function of $n$ for different values of $D$ at $B = 0$ T is plotted in log scale in Fig. S1. Fig. S2(a) shows the $R_{xx}$ as function of $n$ for different values of $D$ at B = 0.7 T. The presence of a small non-quantizing magnetic field enhances the displacement field-dependent asymmetrical feature. In Fig. S2(b) we plot the $R_{xx}$ measured at $n = +1 \times 10^{16}$ m$^{-2}$ and $n = -1 \times 10^{16}$ m$^{-2}$ as a function of D. Here we see for $n = +1 \times 10^{16}$ m$^{-2}$ (blue data points), the resistance remains unchanged for positive values of $D$. On the other hand resistance increases on increasing negative $D$. The behaviour is opposite for $n = -1 \times 10^{16}$ m$^{-2}$ (orange filled circles) with the $R_{xx}$ increasing for $D > 0$ and remaining unchanged for $D < 0$. We also see similar opposite behaviour for the peak value of the resistance at $n_t$ and $n_b$ as a function of $D$. This can be better appreciated in Fig. S2(c) which shows the plot of the peak values of $R_{xx}$ at $n_b$ and $n_t$ as a function of $D/\epsilon_0$.

## S2. LANDAU FAN DIAGRAM AT OTHER VALUE OF D

The Landau fan diagrams for two additional electric fields (0.23 V/nm and -0.42 V/nm) are shown in Fig. S3(a-b). Their qualitative behavior is similar to the data shown in the main text. For high negative $D$, we see a reduction in the visibility of Landau levels. For positive $D$, we see an additional set of Landau fans emerging for $n = -0.75 \times 10^{16}$ m$^{-2}$ with a degeneracy of six. With increasing the $D$, the value of $n_a$ increases (Fig. S4). The sign of $n_a$ changes as sign of D is reversed. This establishes the origin of $n_a$ is related to the layer polarisation.

## S3. VALUE OF BAND GAPS

We measured four-probe resistance as a function of $n$ for different values of temperature (Fig. S6) for $D/\epsilon_0 = 0$ V/nm and $D/\epsilon_0 = 0.18$ V/nm. We find an additional hump emerging at $n = -0.6 \times 10^{16}$ m$^{-2}$ for $D/\epsilon_0 = 0.18$ V/nm (marked by the arrow in Fig. S6(a)). From an Arrhenius fit, we find a gap of $\sim 14.4$ meV. Recall that for this value of $n = -0.6 \times 10^{16}$m$^{-2}$, an extra fan emerges in the Landau fan plot, as shown in Fig.4 (b) of the main manuscript. The values of gaps of the peaks in Fig. S6(a) are listed in Table S5.

## S4. SIGNATURE OF BAND FLATTENING AND ENHANCED ELECTRON-PHONON COUPLING

In Fig. S7, we plot the resistivity $\rho$ versus $T$ measured at $n = n'$ (marked in Fig. S6(a)). The resistivity at $n'$ shows a linear increment with temperature, with large slopes $d\rho/dT \sim$ 8.6 $\Omega$/K and $\sim$ 10.4 $\Omega$/K for $D/\epsilon_0 = 0$ and $D/\epsilon_0 = 0.18$ V/nm, respectively. The value of $d\rho/dT$, which is much larger than the intrinsic monolayer and bilayer graphene [1–3], is attributed to the enhanced electron-phonon coupling arising due to the suppression of Fermi velocity in the reconstructed mini bands [4, 5]. To estimate the Fermi velocity, we use the linear in $T$ resistivity model at high T arising from acoustic phonon-induced scattering, given by[1, 6]

$$\rho(T) = \frac{\pi D_{def}^2}{g_s g_v e^2 \hbar \rho_m v_F^2 v_{ph}^2} k_B T. \quad (1)$$

Here $D_{def}$, $v_{ph}$, $v_F$, and $\rho_m$ are the deformation potential, phonon velocity, Fermi velocity of bilayer graphene, and the atomic mass density, respectively. $g_s = g_v = 2$ are the spin and valley degeneracy, respectively. Using $\rho_m = 7.6 \times 10^{-7}$ kg/m$^2$, $v_{ph} = 2 \times 10^4$ m/s, and $D_{def} \sim 20$ eV [1–3], the estimated Fermi velocity $v_F$ was found to be $1.25 \times 10^5$ m/s and $1.14 \times 10^5$ m/s for $D/\epsilon_0 = 0$ and $D/\epsilon_0 = 0.18$ V/nm, respectively. These values are one order of magnitude smaller than the $V_F$ of pristine BLG. The strong reduction of Fermi velocity with increasing $D$ demonstrates band flattening in our device, in accordance with theoretical calculations [4, 7].

## S5. ADDITIONAL DATA

The contour plot of two-probe resistance (for measurement configuration see inset Fig. S8 (a)) as a function of $n$ and D at 4 K is plotted in Fig. S8 (a). We see qualitatively similar result as seen in four terminal measurement configuration. One can see the asymmetrical feature around the line of charge neutrality as a function of applied D. The data is for another configuration and also for a different cool down (note that the thermal cycle may change the disorder configuration). In Fig. S8 (b), (c) $G$ is plotted as a function of n for different values of positive and negative D respectively. The labels in the plots indicate the value of $D/\epsilon_0$ in units of V/nm.

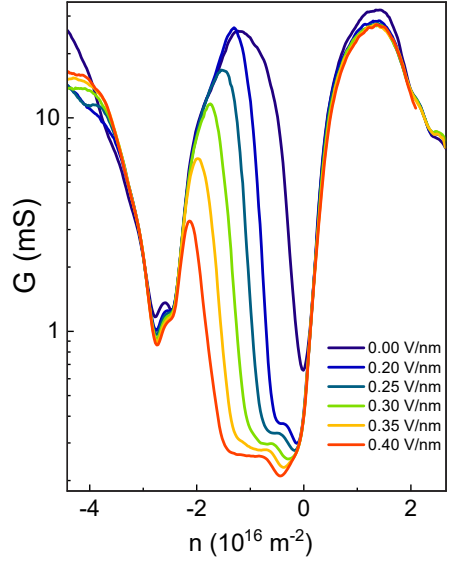

Figure S1. Plot of the longitudinal conductance a function of $n$ for different values of $D/\epsilon_0$ plotted in semi-log scale. The data are for $B = 0$ T.

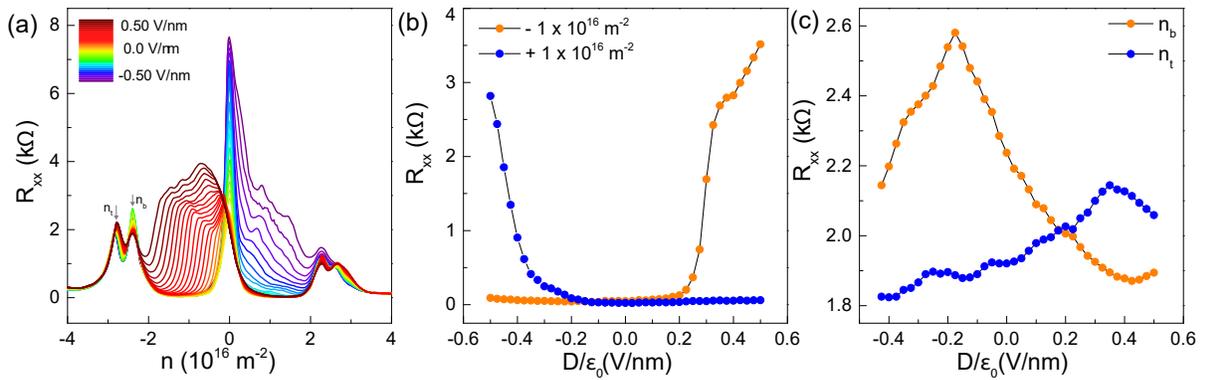

Figure S2. (a) $R_{xx}$ as a function of $n$ for different values of $D/\epsilon_0$, at $B = 0.7$ T. (b) Plot of $R_{xx}$ at $n = +1 \times 10^{16}$ m$^{-2}$ and $n = -1 \times 10^{16}$ m$^{-2}$ as a function of $D/\epsilon_0$. The $R_{xx}$ peak at $n_b$ and $n_t$ are showing completely behaviour. (c) Plot of $R_{xx}$ at $n_b$ and $n_t$ as a function of $D/\epsilon_0$.

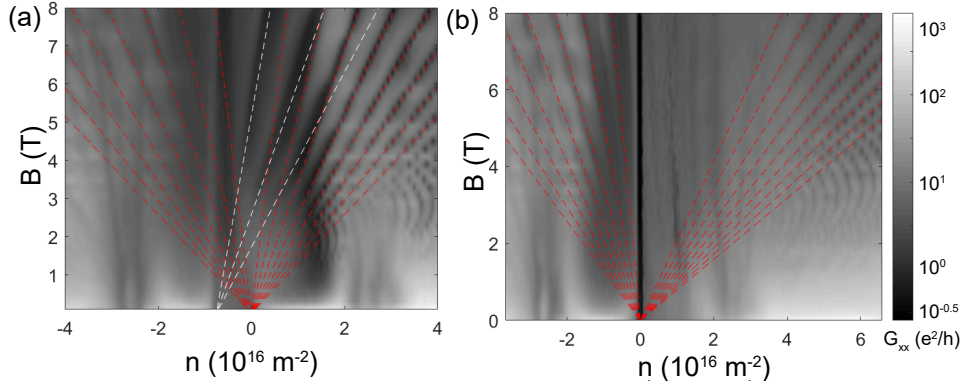

Figure S3. (a) $G_{xx}$ as a function of $n$ and $B$ at $D/\epsilon_0 = 0.23$ V/nm. The red dashed lines indicate the conductance minima at $\nu = \pm 4, \pm 8, \pm 12, \ldots$. An additional landau fan emerges from $n = -0.8 \times 10^{16}$ m$^{-2}$ with a six-fold degeneracy (indicated by white dashed lines). (b) $G_{xx}$ as a function of $n$ and $B$ at $D/\epsilon_0 = -0.42$ V/nm. The red dashed lines indicate the conductance minima at $\nu = \pm 4, \pm 8, \pm 12, \ldots$. For small positive $n$ between $n = 0$ to $n = 2 \times 10^{16}$ m$^{-2}$ the features in $G_{xx}$ is missing.

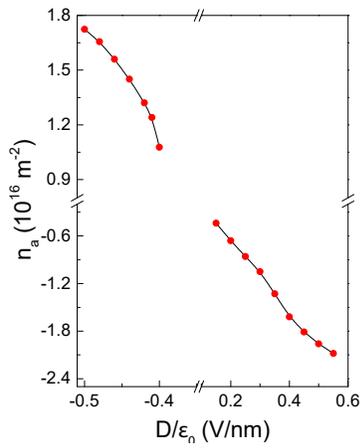

Figure S4. A plot of $n_a$ versus $D/\epsilon_0$. The red dots are the data points; the black line is a guide to the eye.

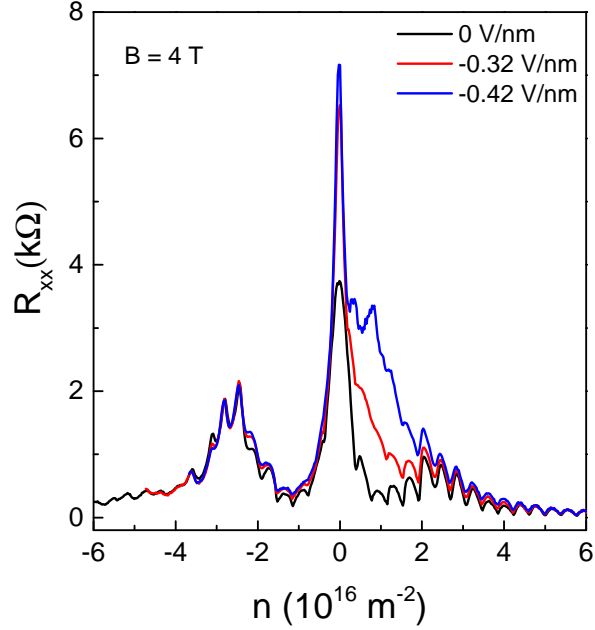

Figure S5. $R_{xx}$ as a function of $n$ at $B = 4$ T for $D/\epsilon_0 = 0, -0.32, -0.42$ V/nm. We see for positive $n$ close to Dirac point, the minima for $R_{xx}$ increases with negative D. This reduces the Landau level visibility.

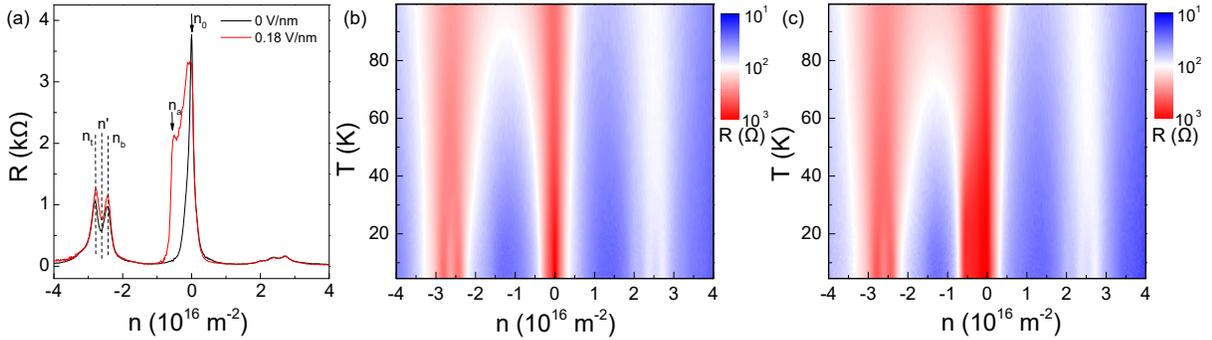

Figure S6. (a) Four-probe resistance as a function of $n$ at $D/\epsilon_0 = 0$ V/nm and $D/\epsilon_0 = 0.18$ V/nm. the different position of peaks are marked. (b) The colour plot of four-probe resistance as a function of $n$ and temperatures T at $D/\epsilon_0 = 0$ V/nm. (c) The colour plot of four-probe resistance as a function of $n$ and T at $D/\epsilon_0 = 0.18$ V/nm.

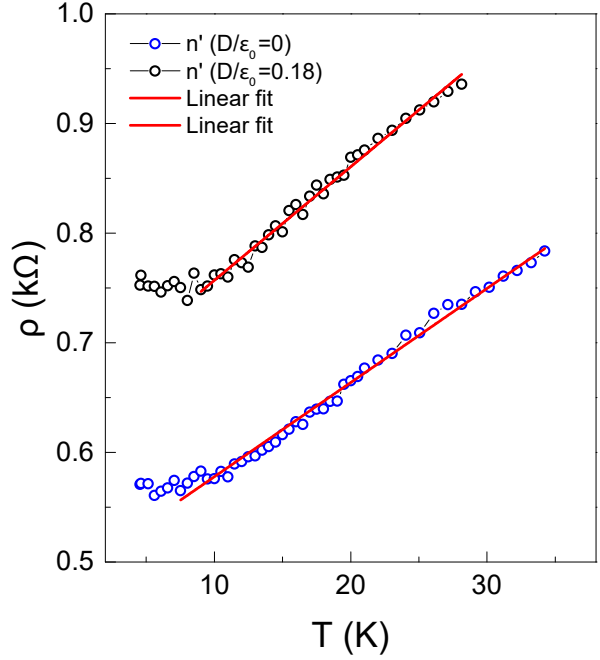

Figure S7. (a) $\rho$ as a function of T at $n = n'$ for $D/\epsilon_0 = 0$ V/nm and $D/\epsilon_0 = 0.18$ V/nm. Red lines are linear fit to the data.

| $n$ | $D/\epsilon_0$ (V/nm) | Energy gap (meV) |
|---|---|---|
| $n_0$ | 0 | 6.4 meV |
| $n_t$ | 0 | 1.03 meV |
| $n_b$ | 0 | 1.3 meV |
| $n_0$ | 0.18 | 11.7 meV |
| $n_t$ | 0.18 | 1.22 meV |
| $n_b$ | 0.18 | 1.5 meV |
| $n_a$ | 0.18 | 14.4 meV |

Table I. Calculated energy gap using plot of Fig. S6. The positions of n is marked in Fig. S6.

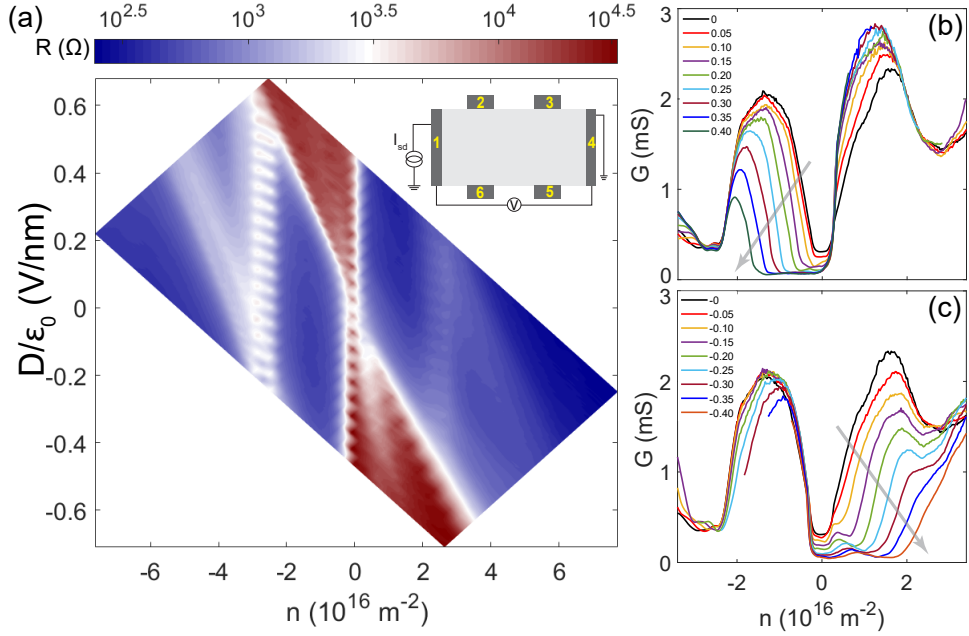

Figure S8. (a) A contour map of two probe resistance $R(B = 0)$ in the $n - D$ plane. (b), (c) $G(B = 0)$ as a function of n for different values of D. The labels in the plots indicate the value of $D/\epsilon_0$ in units of V/nm.

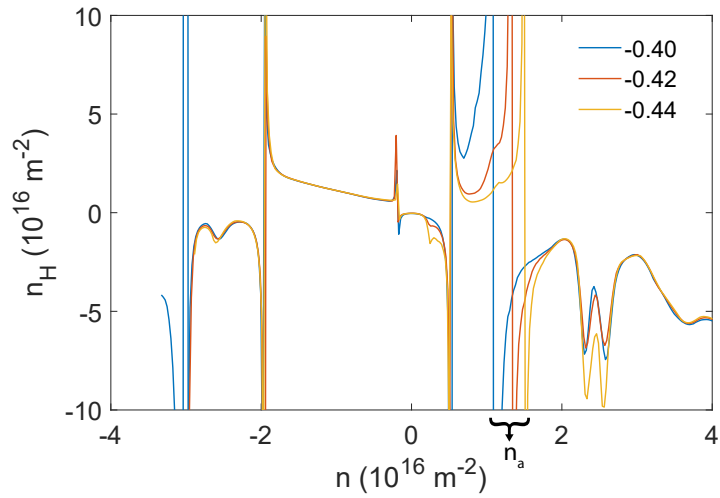

Figure S9. Hall carrier density $n_H$ measured at $B = 0.7$ T as a function of number density $n$ for different values of negative $D/\epsilon_0$ in the units of V/nm.

8

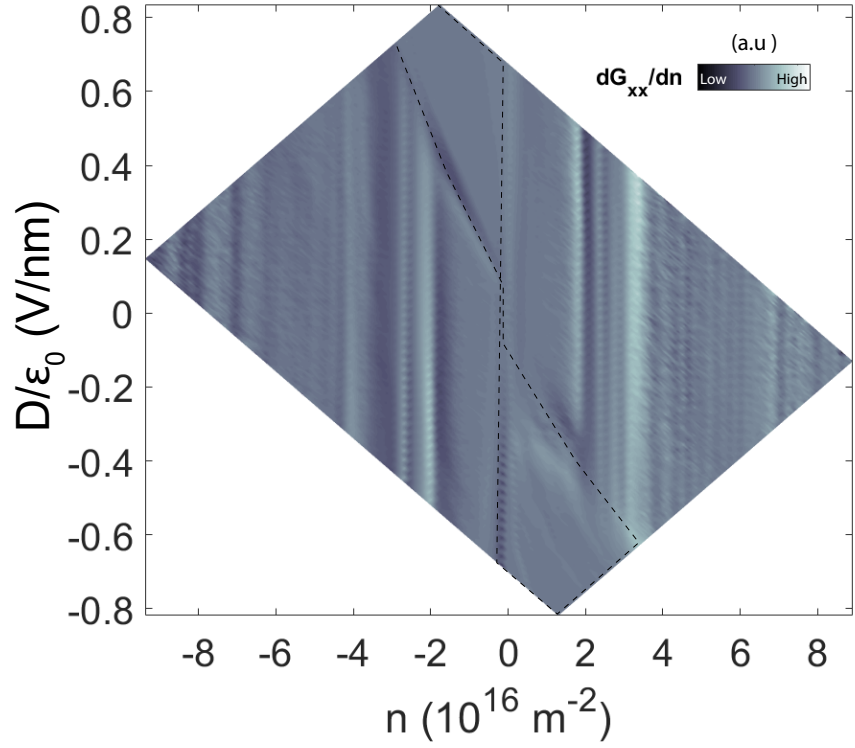

Figure S10. (a) A contour map of $dG_{xx}/dn(B = 0.7\ T)$ in the $n - D$ plane. The asymmetrical feature is marked by black dashed line. The vertical straight features, which are independent of the electric field, are Bragg gaps.

* [priya.tiwari@weizmann.ac.il,](mailto:priya.tiwari@weizmann.ac.il) [aveek@iisc.ac.in](mailto:aveek@iisc.ac.in)


\end{document}